\documentclass[aps,prx,reprint,superscriptaddress,longbibliography]{revtex4-2}
\usepackage{amssymb} 
\usepackage{amsmath}
\usepackage{graphicx}
\usepackage{fancyhdr}
\usepackage{ifthen}
\usepackage{multirow}
\usepackage{footmisc}
\usepackage{array}
\usepackage{color}
\usepackage{ragged2e}
\usepackage{CJKutf8}
\usepackage{hyperref}
\usepackage{physics}

\usepackage{lipsum}
\DeclareSymbolFont{myletters}{OML}{ztmcm}{m}{it}
\DeclareMathSymbol{\uplambda}{\mathord}{myletters}{"15}

\usepackage[]{hyperref}
\usepackage{xcolor}
\definecolor{LinkColor}{RGB}{195,57,57}
\hypersetup{colorlinks=true,citecolor=LinkColor,linkcolor=LinkColor,urlcolor=LinkColor}

\newcounter{aq}
\definecolor{darkgreen}{RGB}{0, 128, 0}

{}

\begin{document}
\title{Phonon-Localization-Driven Decoupling of Dual-Channel Transport for Record-Low Intrinsic Lattice Thermal Conductivity}
\author{Zhunyun Tang}
\date{\today}
\affiliation{School of Physics and Optoelectronics, Xiangtan University, Xiangtan 411105, Hunan, China}
\affiliation{Key Laboratory of Computational Condensed Matter Physics and Materials Quantum Engineering of Hunan Provincial Universities, Xiangtan University, Xiangtan 411105, China}

\author{Xiaoxia Wang }
\affiliation{School of Physics and Optoelectronics, Xiangtan University, Xiangtan 411105, Hunan, China}
\affiliation{Key Laboratory of Computational Condensed Matter Physics and Materials Quantum Engineering of Hunan Provincial Universities, Xiangtan University, Xiangtan 411105, China}

\author{Jin Li}
\affiliation{School of Physics and Optoelectronics, Xiangtan University, Xiangtan 411105, Hunan, China}
\affiliation{Key Laboratory of Computational Condensed Matter Physics and Materials Quantum Engineering of Hunan Provincial Universities, Xiangtan University, Xiangtan 411105, China}

\author{Chaoyu He}
\affiliation{School of Physics and Optoelectronics, Xiangtan University, Xiangtan 411105, Hunan, China}
\affiliation{Key Laboratory of Computational Condensed Matter Physics and Materials Quantum Engineering of Hunan Provincial Universities, Xiangtan University, Xiangtan 411105, China}

\author{Chao Tang}
\thanks{\textnormal{} 
	\href{mailto:tang\_chao@xtu.edu.cn}{tang\_chao@xtu.edu.cn}.}
\affiliation{School of Physics and Optoelectronics, Xiangtan University, Xiangtan 411105, Hunan, China}
\affiliation{Key Laboratory of Computational Condensed Matter Physics and Materials Quantum Engineering of Hunan Provincial Universities, Xiangtan University, Xiangtan 411105, China}

\author{Mingxing Chen}
\thanks{\textnormal{}
\href{mailto:mxchen@hunnu.edu.cn}{mxchen@hunnu.edu.cn}.}
\affiliation{School of Physics and Electronics, Hunan Normal University, Changsha 410081, Hunan, China}

\author{Tao Ouyang}
\thanks{\textnormal{} 
\href{mailto:ouyangtao@xtu.edu.cn}{ouyangtao@xtu.edu.cn}.}
\affiliation{School of Physics and Optoelectronics, Xiangtan University, Xiangtan 411105, Hunan, China}
\affiliation{Key Laboratory of Computational Condensed Matter Physics and Materials Quantum Engineering of Hunan Provincial Universities, Xiangtan University, Xiangtan 411105, China}

\begin{abstract}
A fundamental bottleneck in pushing the intrinsic lattice thermal conductivity of inorganic crystalline solids to its lowest limit arises from the inherent competition between the particle-like propagation (\(\kappa_{\mathrm{L}}^{\mathrm{P}}\)) and wave-like tunneling (\(\kappa_{\mathrm{L}}^{\mathrm{C}}\)) channels. Herein, we demonstrate that phonon localization provides a robust pathway to decouple the dual-channel transport, achieving record-low \(\kappa_{\mathrm{L}}\) in quasi-1D ternary helical crystals. Despite the structural complexity leading to densely populated phonon branches and thus inducing abundant coherent phonons, the weak interchain interactions and heavy elements compress numerous branches into highly localized, nearly dispersionless flat bands. Such strong localization simultaneously suppresses both the diagonal and off-diagonal components of the group velocity, thereby synergistically suppressing \(\kappa_{\mathrm{L}}^{\mathrm{P}}\) and \(\kappa_{\mathrm{L}}^{\mathrm{C}}\). Taking InSeI as an example, the interchain room-temperature \(\kappa_{\mathrm{L}}^{\mathrm{P}}\) and \(\kappa_{\mathrm{L}}^{\mathrm{C}}\) are 0.145 and 0.053 W/mK, respectively, yielding an ultralow total \(\kappa_{\mathrm{L}}\) of 0.198 W/mK. Weaker interchain interactions further drive the room-temperature \(\kappa_{\mathrm{L}}\) of GaSeI and AlSeI to record lows of 0.086 and 0.089 W/mK, respectively; these values even drop to 0.058 and 0.059 W/mK at 900 K. These findings provide useful insights into exploring the thermal conductivity limit in crystals. 
\end{abstract}

\maketitle



\section{Introduction}\label{sec:1}	
Exploring inorganic crystalline solids with intrinsically ultralow lattice thermal conductivity (\(\kappa_{\mathrm{L}}\)) is crucial for both fundamental condensed matter physics and a wide range of technological applications, including high-performance thermoelectrics, thermal barrier coatings, and advanced thermal management in microelectronic devices~\cite{RN1,RN2,RN3,RN4,RN5}. Since lattice vibrations (phonons) are the primary heat carriers in electrically insulating materials, suppressing phonon-mediated heat transport has been the central focus of materials design. Over the past decade, significant progress has been achieved on both theoretical and experimental fronts by leveraging a variety of intrinsic phonon scattering mechanisms, including complex crystal structures~\cite{RN6,RN7}, stereochemically active lone-pair electrons~\cite{RN8,RN9,RN10}, rattling guest atoms in oversized cages~\cite{RN11,RN12,RN13}, and bonding heterogeneity or antibonding states~\cite{RN14,RN15,RN16,RN17,RN18,RN19}. These design principles have led to the discovery of numerous crystalline thermal insulators with \(\kappa_{\mathrm{L}}\) values approaching the amorphous limit. For example, \(\mathrm{Ba_{8}Ga_{16}Ge_{30}}\)~\cite{RN7}, \(\mathrm{AgInSnSe_{4}}\)~\cite{RN10}, \(\mathrm{CuBiI_{4}}\)~\cite{RN19}, \(\mathrm{CsCu_{2}I_{3}}\)~\cite{RN20}, and \(\mathrm{Cs_{3}Bi_{2}Br_{9}}\)~\cite{RN21} exhibit extremely low \(\kappa_{\mathrm{L}}\) values of 0.97, 0.44, 0.38, 0.3, and 0.25 W/mK at room temperature, respectively.

Despite these successes, recent advances in thermal transport theory have uncovered a fundamental bottleneck in pushing \(\kappa_{\mathrm{L}}\) to its ultimate minimum. In the unified theory of thermal transport developed by Simoncelli et al.~\cite{RN22,RN23}, \(\kappa_{\mathrm{L}}\) is rigorously decomposed into two distinct channels: (i) the particle-like contribution (\(\kappa_{\mathrm{L}}^{\mathrm{P}}\)) arising from semi-classical phonon propagation, and (ii) the wave-like tunneling contribution (\(\kappa_{\mathrm{L}}^{\mathrm{C}}\)) originating from interband quantum coherence. For strong anharmonic materials with structurally complex unit cells, although the particle-like channel (\(\kappa_{\mathrm{L}}^{\mathrm{P}}\)) can be remarkably suppressed, the dense distribution of phonon branches inevitably facilitates pronounced wave-like tunneling (\(\kappa_{\mathrm{L}}^{\mathrm{C}}\)) due to the enhanced coherence between adjacent near-degenerate phonon modes. This inherent interplay often diminishes the overall reduction in \(\kappa_{\mathrm{L}}\), manifesting as temperature-independent or even positive temperature-dependent behavior~\cite{RN24,RN25,RN26,RN27,RN28,RN29,RN30}. For instance, for \(\mathrm{Cu_{7}PS_{6}}\) and \(\mathrm{Ag_{8}SnS_{6}}\), despite the particle-like \(\kappa_{\mathrm{L}}^{\mathrm{P}}\) being nearly as low as that of air (\(\sim 0.04\) W/mK), the resulting strong wave-like \(\kappa_{\mathrm{L}}^{\mathrm{C}}\) exceeds 0.3 W/mK, severely hindering the reduction of the total \(\kappa_{\mathrm{L}}\)~\cite{RN24,RN29}. Consequently, the intrinsic trade-off between \(\kappa_{\mathrm{L}}^{\mathrm{P}}\) and \(\kappa_{\mathrm{L}}^{\mathrm{C}}\) poses a formidable challenge to decouple the dual-channel thermal transport in crystalline solids.

\begin{figure*}[t!] 
	\centering
	\includegraphics[width=1.0\linewidth]{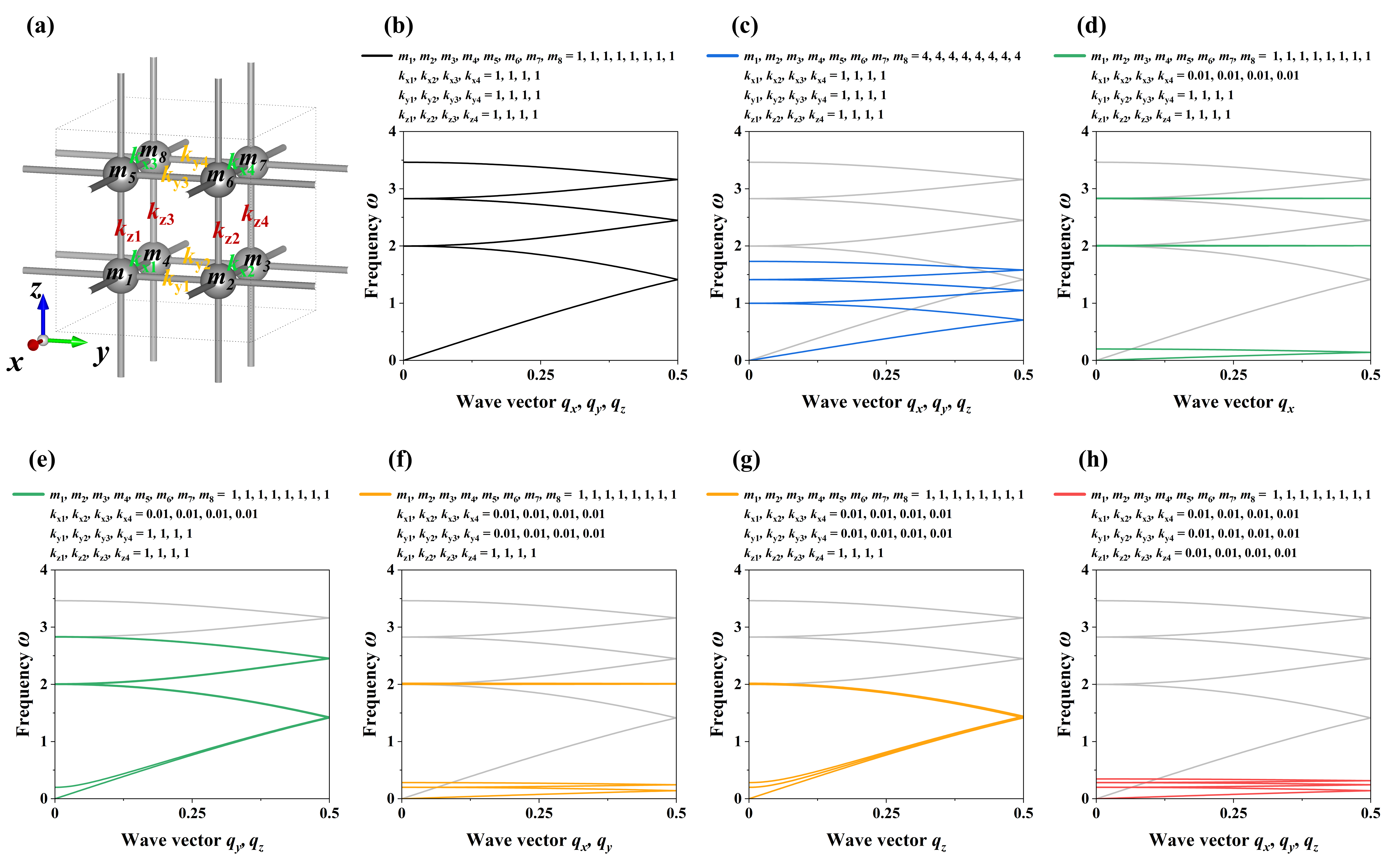}
	\caption{(a) Schematic diagram of a simplified eight-atom lattice model. (b–h) Phonon dispersion curves along different wave-vector directions \((q_{x}, q_{y}, q_{z})\), illustrating the evolution of phonon branches as atomic mass (\(m\)) and bond strength (\(k\)) vary.}
	\label{fgr:fig-1}
\end{figure*}

To overcome this dilemma, researchers primarily focused on exploring materials that exhibit both strong anharmonicity and simple crystal structures. In this ideal scenario, the pronounced anharmonicity largely inhibits particle-like phonon transport. Meanwhile, the simple lattice gives rise to sparse phonon branches, rendering the wave-like tunneling effect comparatively weak. Representative examples of this strategy include the simple crystalline systems \(\mathrm{Tl_{3}VSe_{4}}\) (\(\kappa_{\mathrm{L}}\approx 0.30\) W/mK)~\cite{RN31}, \(\mathrm{AgTlI_{2}}\) (\(\kappa_{\mathrm{L}}\approx 0.25\) W/mK)~\cite{RN32}, and \(\mathrm{AgTl_{2}I_{3}}\) (\(\kappa_{\mathrm{L}}\approx 0.21\) W/mK)~\cite{RN18}. However, compared with structurally complex systems, simple structures still lack the natural advantage of achieving strong anharmonic phonon scattering, making it difficult to push their \(\kappa_{\mathrm{L}}^{\mathrm{P}}\) to the ultralow limit. To date, the lower limit of \(\kappa_{\mathrm{L}}\) in crystalline materials remains generally constrained to the 0.1–0.3 W/mK range. In other words, it remains challenging to simultaneously suppress both particle-like and wave-like thermal transport from the perspective of anharmonic scattering.

Based on the above insights, we propose that an alternative pathway for synergistically suppressing both \(\kappa_{\mathrm{L}}^{\mathrm{P}}\) and \(\kappa_{\mathrm{L}}^{\mathrm{C}}\) lies in driving the lattice toward a strong phonon localization regime. This limit generally requires nearly dispersionless phonon branches. Under this condition, the group velocity matrix elements approach zero, thereby fundamentally achieving synergistic suppression of dual-channel thermal transport. Such flattened phonon bands could be achieved by introducing heavy elements and weakening interatomic bonding interactions~\cite{RN33,RN34}. To clearly reveal the impact of atomic mass and interatomic interactions on the phonon spectrum, we construct an eight-atom model, as illustrated in Fig.~\ref{fgr:fig-1}(a). By comparing Figs.~\ref{fgr:fig-1}(b) and \ref{fgr:fig-1}(c), it can be observed that the phonon branches become flatter as the mass parameter increases. Moreover, we simulate quasi-low-dimensional lattice dynamics by significantly weakening the interaction strength along specific directions. When the bonding strength along the \(x\)-direction is significantly weaker than that along the other two directions, the system evolves into a quasi-two-dimensional model [Figs.~\ref{fgr:fig-1}(d–e)]. Simultaneously weakening the interactions along both the \(x\)- and \(y\)-directions can further tune the system to the quasi-one-dimensional limit [Figs.~\ref{fgr:fig-1}(f–g)]. When the coupling strengths in all three directions are reduced to extremely low values [Fig.~\ref{fgr:fig-1}(h)], the system eventually evolves into a quasi-zero-dimensional model. Notably, lattice vibrations along the quasi-low-dimensional direction are strongly confined, resulting in phonon branches that exhibit highly localized and nearly dispersionless behavior [Figs.~\ref{fgr:fig-1}(d), \ref{fgr:fig-1}(f), and \ref{fgr:fig-1}(h)]. Following these design principles, the recently synthesized family of quasi-one-dimensional (quasi-1D) van der Waals (vdW) III–VI–VII ternary helical crystals (including InSeI, GaSI, GaSeI, and AlSeI) appears to possess unique advantages~\cite{RN35,RN36,RN37,RN38,RN39}. These materials exhibit fascinating structural chirality, wide bandgaps, and robust non-linear optical properties. More importantly, they consist of 1D tetrahelical-tube chains characterized by weak interchain vdW interactions, a high proportion of heavy elements, and a complex atomic environment. These features render them an ideal platform for investigating the decoupling of dual-channel thermal transport.

Herein, we systematically investigate the lattice dynamics and thermal transport properties of this unique class of quasi-1D helical materials by combining first-principles calculations, machine learning potentials, molecular dynamics simulations, and the unified thermal transport theory. The results reveal strong phonon localization in these complex quasi-1D structures, leading to extremely low phonon group velocities, which are significantly lower than those of previously reported materials with ultralow \(\kappa_{\mathrm{L}}\). Taking the representative compound InSeI as an example, it is observed that both the particle-like propagation and wave-like tunneling channels are simultaneously and strongly suppressed. Specifically, at room temperature, the interchain thermal conductivity components of \(\kappa_{\mathrm{L}}^{\mathrm{P}}\) and \(\kappa_{\mathrm{L}}^{\mathrm{C}}\) are suppressed to 0.145 and 0.053 W/mK, respectively, leading to an ultralow total \(\kappa_{\mathrm{L}}\) of only 0.198 W/mK. More remarkably, GaSeI exhibits a lower interchain \(\kappa_{\mathrm{L}}\) of 0.086 W/mK at 300 K, establishing a new record for inorganic crystalline thermal insulators. These findings explicitly demonstrate that strong phonon localization provides a powerful route to decouple the dual-channel transport, offering a new paradigm for designing crystalline materials with ultralow \(\kappa_{\mathrm{L}}\) via van der Waals dimensional confinement and phonon engineering.

\section{Results and Discussion}\label{sec:2}

\begin{figure*}[t!] 
	\centering
	\includegraphics[width=1.0\linewidth]{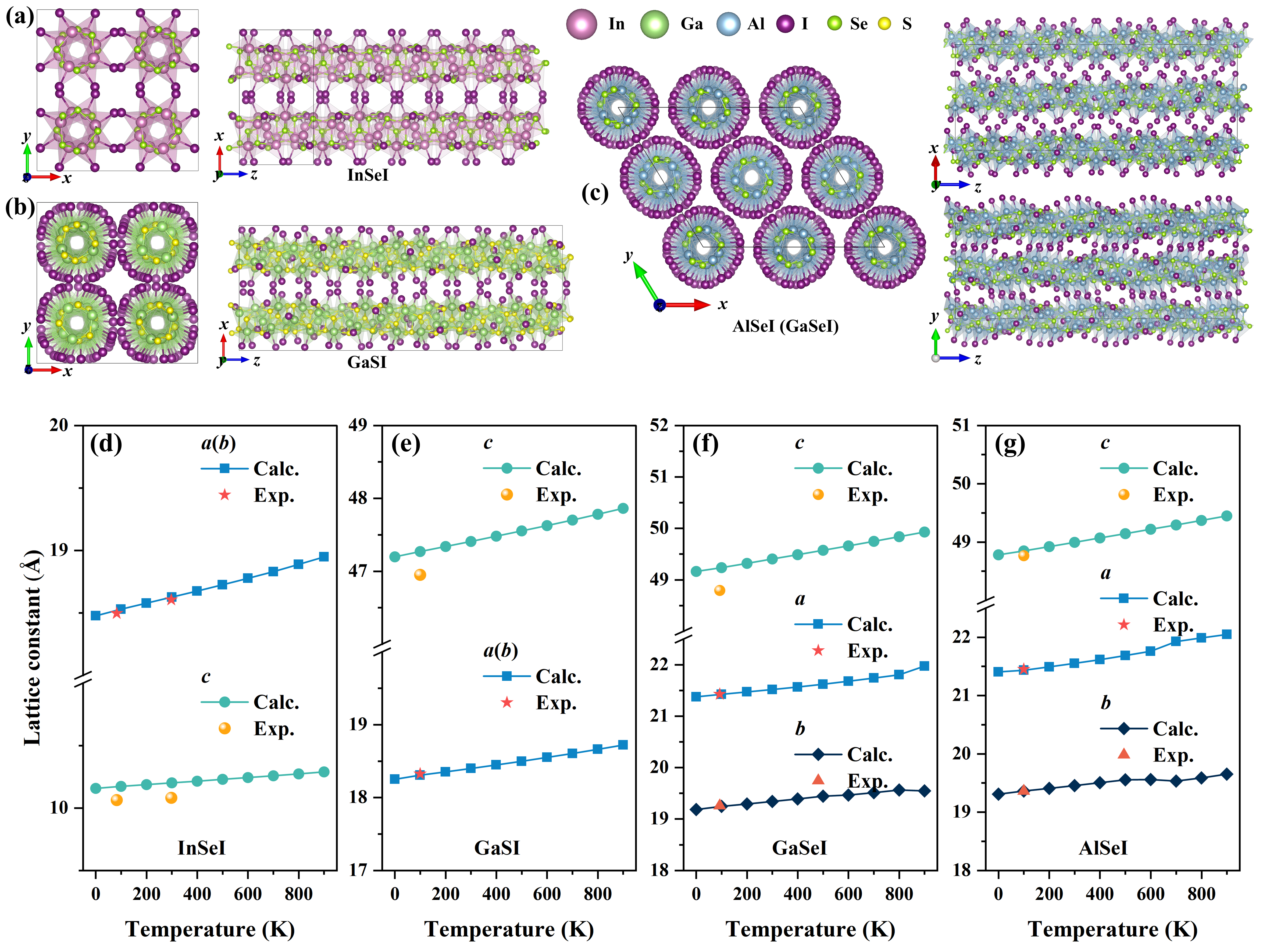}
	\caption{(a–c) Crystal structures of InSeI, GaSI, and isostructural GaSeI/AlSeI, viewed in different projections. (d–g) Comparison of simulated and experimental lattice constants for the four compounds at different temperatures.}
	\label{fgr:fig-2}
\end{figure*}
\begin{figure*}[t!] 
	\centering
	\includegraphics[width=1.0\linewidth]{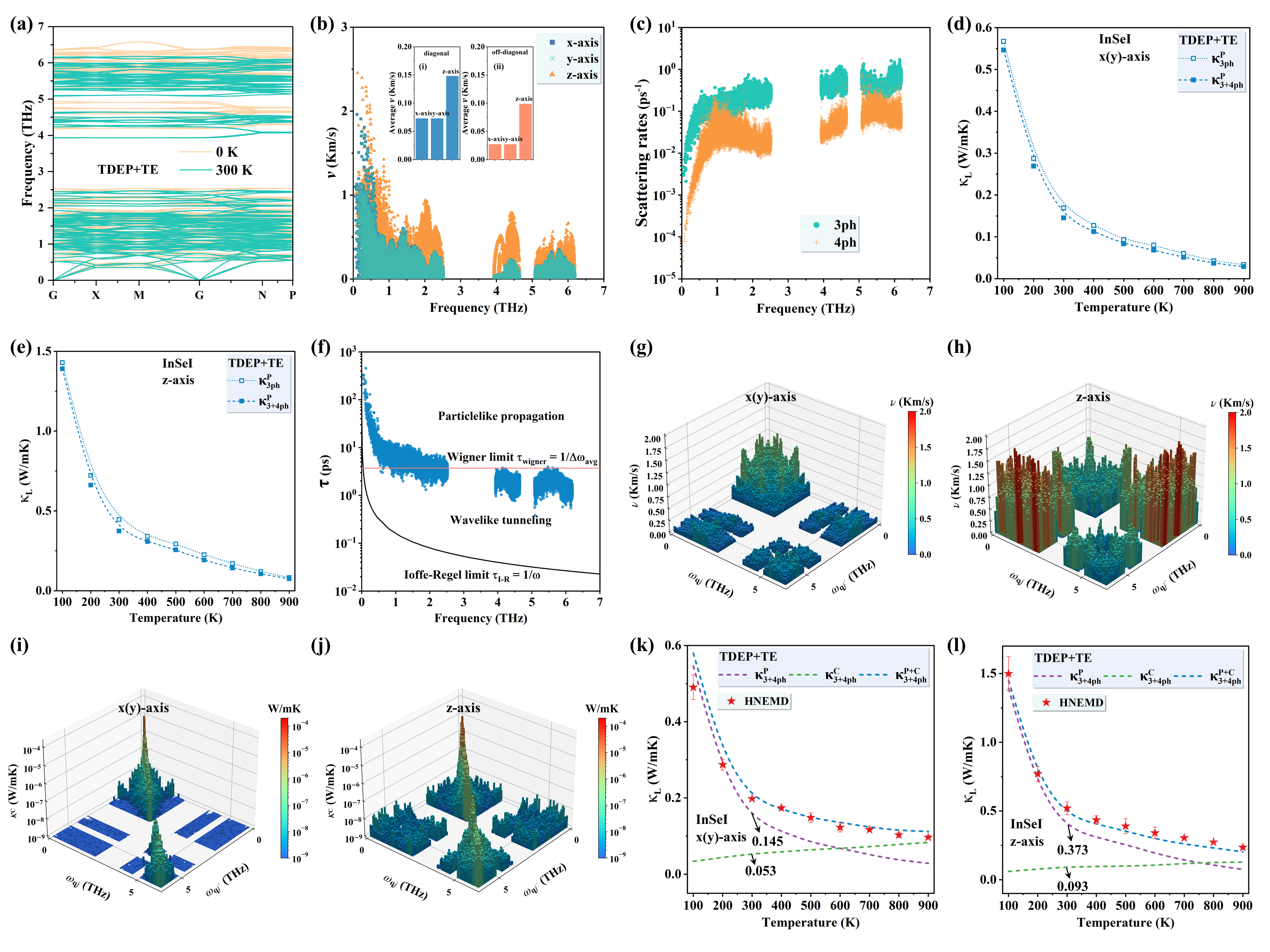}
	\caption{(a) Phonon dispersion curves of InSeI at 0 and 300 K, obtained via temperature-dependent effective potential with thermal expansion (TDEP+TE) considered. (b) Diagonal phonon group velocity (\(v\)) along different directions at 300 K. (c) Three- (3ph) and four-phonon (4ph) scattering rates at 300 K. (d–e) Particle-like thermal conductivity (\(\kappa_{\mathrm{L}}^{\mathrm{P}}\)) along different directions as a function of temperature, calculated with only 3ph scattering (\(\kappa_{3\mathrm{ph}}^{\mathrm{P}}\)) and with both 3ph and 4ph scattering (\(\kappa_{3+4\mathrm{ph}}^{\mathrm{P}}\)). (f) Phonon lifetime at 300 K. The red and black lines denote the Wigner and Ioffe–Regel limits, respectively. Here, \(\omega = 2\pi f\), where \(f\) is the phonon frequency in THz. \(\Delta \omega_{\mathrm{avg}} = \omega_{\mathrm{max}} / 3N_{\mathrm{at}}\), where \(\Delta \omega_{\mathrm{avg}}\), \(N_{\mathrm{at}}\), and \(\omega_{\mathrm{max}}\) denote the average interband spacing, the number of atoms in the primitive cell, and the maximum phonon frequency, respectively. Three-dimensional visualization of mode-resolved (g–h) off-diagonal \(v\) and (i–j) coherent thermal conductivity (\(\kappa_{\mathrm{L}}^{\mathrm{C}}\)) along different directions at 300 K. (k–l) \(\kappa_{3+4\mathrm{ph}}^{\mathrm{P}}\), \(\kappa_{3+4\mathrm{ph}}^{\mathrm{C}}\), and \(\kappa_{3+4\mathrm{ph}}^{\mathrm{P+C}}\) along different directions as a function of temperature, validated by homogeneous nonequilibrium molecular dynamics (HNEMD) simulations. Insets (i) and (ii) in (b) show the average \(v\) values for diagonal and off-diagonal terms, respectively.}
	\label{fgr:fig-3}
\end{figure*}

The quasi-one-dimensional (quasi-1D) III–VI–VII ternary helical crystals investigated in this work crystallize in distinct yet structurally related lattice types, all characterized by 1D van der Waals (vdW) chains built from corner-sharing [TX\(_3\)Y] tetrahedral units (where \(\mathrm{T} = \mathrm{In}\), Ga, Al; \(\mathrm{X} =\) Se, S; \(\mathrm{Y} = \mathrm{I}\)). As shown in Fig.~\ref{fgr:fig-2}(a), InSeI crystallizes in the tetragonal space group \(I4_{1}/a\), featuring a periodic \(4_{1}\) screw axis where the helical chains are packed in a simple tetragonal motif. The tetrahedra [InSe\(_3\)I] are bridged by Se atoms, propagating the helix along the \(z\)-axis. Substituting Se with S and In with Ga yields GaSI, which crystallizes in the non-centrosymmetric tetragonal \(\mathrm{P\bar{4}}\) space group [Fig.~\ref{fgr:fig-2}(b)]. Despite sharing the tetragonal chain packing, GaSI manifests a unique “squircular” helix cross-section due to the smaller atomic radius of S, which induces an unusual angular strain within the inner tubule of the chain. In contrast, both GaSeI and AlSeI, which are isostructural, adopt the monoclinic \(Cc\) space group. In these phases, the helical chains are arranged in a pseudo-hexagonal stacking pattern, with the central T atoms forming near-ideal Boerdijk–Coxeter tetrahelices characterized by an irrational twist angle [Fig.~\ref{fgr:fig-2}(c)]. All these unique 1D chains are held together by weak interchain vdW forces, which fundamentally govern their anisotropic dynamic behavior. In Figs.~\ref{fgr:fig-2}(d–g), we first simulate their thermal expansion behavior. For all four compounds, the calculated lattice constants exhibit excellent agreement with the experimental measurements~\cite{RN35,RN36,RN37,RN38,RN39}. Meanwhile, it can be noted that the thermal expansion along the crystallographic \(z\)-axis is distinct from the \(x(y)\)-axis. For instance, in InSeI, the lattice constants \(a(b)\) [along the \(x(y)\)-axis] and \(c\) (along the \(z\)-axis) increase from 18.476 Å and 10.157 Å at 0 K to 18.948 Å and 10.289 Å at 900 K, corresponding to relative expansions of approximately \(2.5\%\) and \(1.3\%\), respectively. A similar trend is observed for the other three compounds. This indicates that thermal expansion along the interchain direction exhibits a stronger temperature dependence than that along the intrachain covalent direction, reflecting the significant thermal softening of the weak interchain vdW bonding.

We then investigate the phonon and thermal transport properties of InSeI. The phonon dispersions are calculated based on the finite-temperature lattice constants using the temperature-dependent effective potential (TDEP) method~\cite{RN40}. As shown in Fig. S5 of the Supporting Information (SI), the phonon branches are quite dense due to the structural complexity and exhibit significant softening with increasing temperature. Furthermore, consistent with the above inference, the presence of numerous heavy elements severely constrains the lattice vibrations, limiting the highest optical branch frequency to below 7 THz and the acoustic branch to below 1 THz. Most phonon modes are localized within a narrow energy range. This strong phonon localization behavior is more clearly visualized in Fig.~\ref{fgr:fig-3}(a), where numerous phonon branches remain nearly dispersionless. Consequently, the phonon group velocity (\(v\)) of InSeI is dramatically suppressed, particularly along the interchain directions governed by weak vdW interactions. As shown in Fig.~\ref{fgr:fig-3}(b), at room temperature, the diagonal group velocities along the interchain direction are generally less than 0.5 km/s within the frequency range above 1 THz. Even along the intrachain (\(z\)-axis) direction, the \(v\) is strictly restricted to less than 1 km/s. The inset (i) of Fig.~\ref{fgr:fig-3}(b) reveals that the average diagonal \(v\) along the interchain direction is as low as 0.073 km/s, which is nearly half of the value along the intrachain direction (0.148 km/s). Obviously, the phonon \(v\) of InSeI is significantly lower than that of other known ultralow \(\kappa_{\mathrm{L}}\) materials~\cite{RN16,RN41,RN42,RN43,RN44,RN45,RN46}.

As illustrated in Fig.~\ref{fgr:fig-3}(c), the phonon scattering rates are further calculated. In InSeI, the scattering process is primarily dominated by three-phonon (3ph) interactions, with only a few modes exhibiting four-phonon (4ph) scattering rates comparable to the 3ph processes. By incorporating these scattering mechanisms, the particle-like lattice thermal conductivity (\(\kappa_{\mathrm{L}}^{\mathrm{P}}\)) is evaluated and presented in Figs.~\ref{fgr:fig-3}(d–e). The results indicate that the thermal transport properties of this system are significantly suppressed. When only 3ph interactions are considered, \(\kappa_{3\mathrm{ph}}^{\mathrm{P}}\) along the interchain and intrachain directions are 0.169 and 0.446 W/mK, respectively. Upon including 4ph scattering, \(\kappa_{3+4\mathrm{ph}}^{\mathrm{P}}\) slightly decrease to 0.145 and 0.373 W/mK, respectively. To elucidate the role of coherent transport, we plot the phonon lifetimes as a function of frequency and mark the Wigner limit~\cite{RN22,RN23} in Fig.~\ref{fgr:fig-3}(f). It is observed that most phonon modes in the high-frequency region lie below this limit, implying that they may exhibit significant wave-like tunneling behavior. However, as illustrated in Figs.~\ref{fgr:fig-3}(g–h), the off-diagonal group velocities along the interchain direction are extremely low and substantially smaller than those along the intrachain direction. The corresponding average values, shown in inset (ii) of Fig.~\ref{fgr:fig-3}(b), are 0.027 and 0.099 km/s along the interchain and intrachain directions, respectively. This suggests that, despite the existence of abundant coherent phonons, their tunneling contribution to thermal transport may be severely hindered by their intrinsically ultralow off-diagonal velocities. This is intuitively confirmed by the three-dimensional visualization of the mode-resolved coherent thermal conductivity presented in Figs.~\ref{fgr:fig-3}(i–j).

Finally, we present the temperature-dependent particle-like (\(\kappa_{3+4\mathrm{ph}}^{\mathrm{P}}\)), coherent (\(\kappa_{3+4\mathrm{ph}}^{\mathrm{C}}\)), and total (\(\kappa_{3+4\mathrm{ph}}^{\mathrm{P+C}}\)) thermal conductivities in Figs.~\ref{fgr:fig-3}(k–l). These results clearly demonstrate that the decoupling of dual-channel transport is successfully achieved in InSeI. At room temperature, the \(\kappa_{3+4\mathrm{ph}}^{\mathrm{P}}\) and \(\kappa_{3+4\mathrm{ph}}^{\mathrm{C}}\) along the interchain direction are 0.145 and 0.053 W/mK, respectively, resulting in an ultralow \(\kappa_{3+4\mathrm{ph}}^{\mathrm{P+C}}\) value below 0.2 W/mK. As the temperature rises, the coherent contribution becomes more pronounced but does not fully dominate the thermal transport; therefore, the total thermal conductivity still exhibits a decreasing trend across a wide temperature range. Notably, the thermal conductivity values obtained by using the homogeneous nonequilibrium molecular dynamics (HNEMD) method~\cite{RN47} align well with the \(\kappa_{3+4\mathrm{ph}}^{\mathrm{P+C}}\) values, further confirming the reliability of these results. Furthermore, although the thermal conductivity along the intrachain direction is higher than that along the interchain direction, it still exhibits a relatively low \(\kappa_{3+4\mathrm{ph}}^{\mathrm{P+C}}\) value of 0.466 W/mK at room temperature.

\begin{figure*}[t!] 
	\centering
	\includegraphics[width=1.0\linewidth]{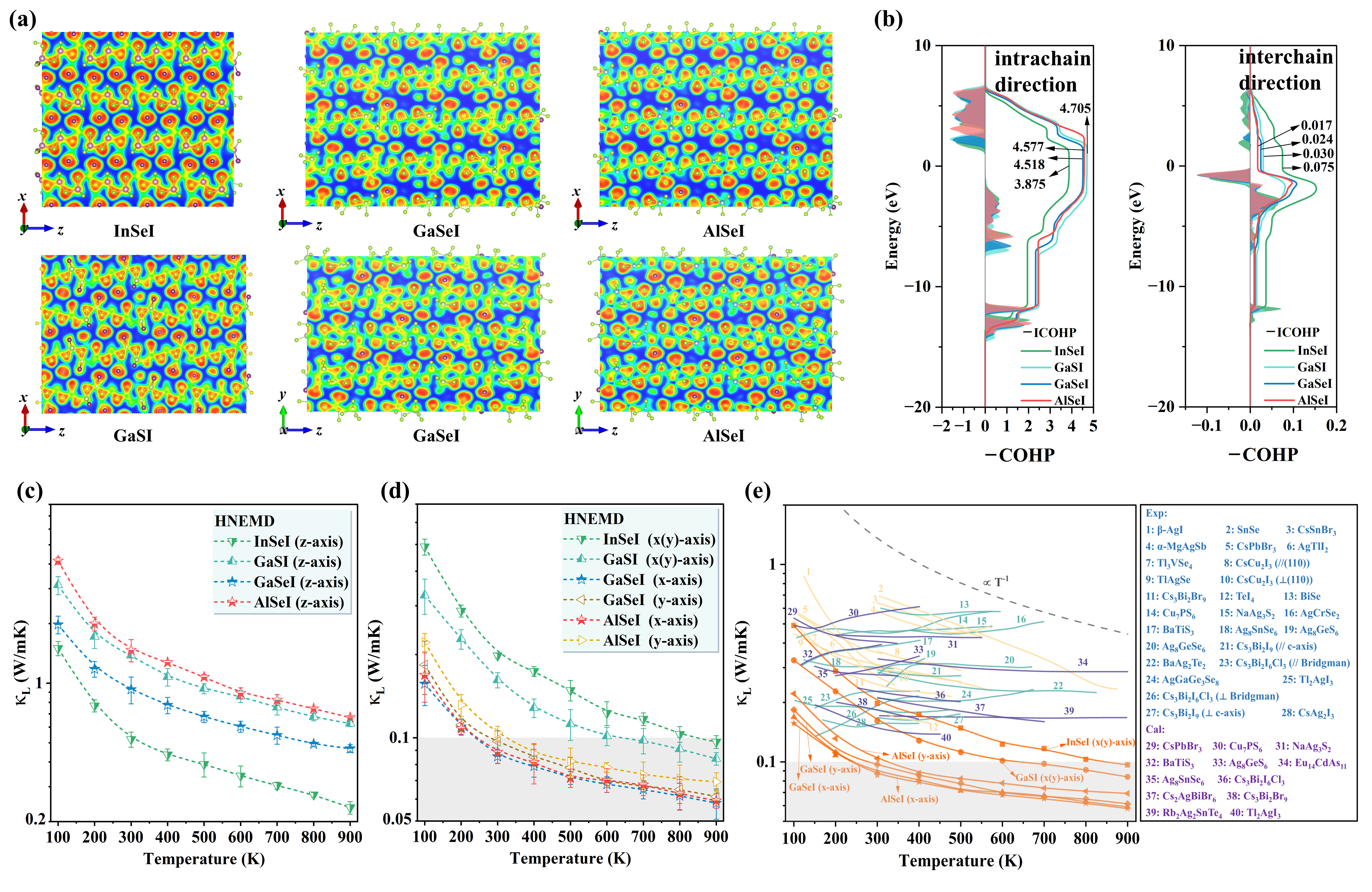}
	\caption{(a) Electron localization function (ELF) of InSeI, GaSI, GaSeI, and AlSeI, viewed in different projections. (b) Average negative crystal orbital Hamilton population (–COHP) and integrated COHP (–ICOHP) along intrachain and interchain directions. The intrachain average is calculated by considering all intrachain bonding atom pairs. The inter-chain average is taken over all pairs formed by each I atom and its neighboring atoms on adjacent chains within 5 Å. (c–d) Intrachain and interchain \(\kappa_{\mathrm{L}}\) as a function of temperature from HNEMD simulations. (e) Comparison of the interchain \(\kappa_{\mathrm{L}}\) of four quasi-1D materials with other typical ultralow \(\kappa_{\mathrm{L}}\) materials.}
	\label{fgr:fig-4}
\end{figure*}

The electron localization function (ELF) is calculated to gain deeper insight into the underlying mechanism governing thermal transport in these quasi-1D materials. As depicted in Fig.~\ref{fgr:fig-4}(a), the ELF is highly localized within the intrachain bonds, implying strong covalent bonding along the helical tubes. In contrast, the ELF is completely delocalized in the interchain regions, confirming that the interactions between chains are primarily governed by weak vdW forces. Notably, compared with InSeI, the other three compounds exhibit more extended delocalization regions along the interchain direction, especially for GaSeI and AlSeI. This phenomenon is mainly attributable to the looser interchain packing observed in those two compounds, as illustrated in Fig.~\ref{fgr:fig-2}(a–c). To further quantify the interatomic interactions, we calculate the integrated crystal orbital Hamilton population (ICOHP)~\cite{RN48} for key atomic pairs along both the intrachain and interchain directions. Figure~\ref{fgr:fig-4}(b) reveals that the intrachain –ICOHP values are nearly two orders of magnitude larger than their interchain counterparts, further confirming the remarkably weak interchain interactions. Specifically, the average intrachain –ICOHP values for InSeI, GaSI, GaSeI, and AlSeI are 3.875, 4.705, 4.518, and 4.577, respectively. Among them, InSeI exhibits the weakest intrachain bonding interactions, consistent with the general trend that heavier atomic systems display weaker bonding~\cite{RN49}. Along the interchain direction, however, the –ICOHP values for GaSI, GaSeI, and AlSeI are 0.030, 0.024, and 0.017, respectively, which are noticeably smaller than that of InSeI (0.075). This behavior is in stark contrast to the intrachain case. Consequently, the \(\kappa_{\mathrm{L}}\) along the intrachain and interchain directions exhibits distinctly different mass dependencies. It is clear from Fig.~\ref{fgr:fig-4}(c) that the intrachain \(\kappa_{\mathrm{L}}\) of the lighter GaSI, GaSeI, and AlSeI compounds are significantly larger than that of InSeI. However, along the interchain directions, an inverse trend is observed, with the lighter systems exhibiting lower \(\kappa_{\mathrm{L}}\) [Fig.~\ref{fgr:fig-4}(d)]. To contextualize the exceptional thermal insulating performance, we benchmark the temperature-dependent interchain \(\kappa_{\mathrm{L}}\) of these quasi-1D materials against previously reported typical ultralow \(\kappa_{\mathrm{L}}\) crystalline materials in Fig.~\ref{fgr:fig-4}(e)~\cite{RN21,RN22,RN23,RN24,RN25,RN26,RN27,RN28,RN29,RN30,RN32,RN41,RN50,RN51,RN52,RN53,RN54,RN55,RN56,RN57,RN58,RN59,RN60,RN61,RN62,RN63,RN64,RN65,RN66,RN67}. In these established thermal insulators, heat transport is dominated by coherent phonons. Consequently, their thermal conductivity exhibits a weak or even positive temperature dependence, which substantially restricts further suppression of thermal transport at high temperatures. In contrast, these quasi-1D materials calculated in this work exhibit a notably steeper and persistent decrease with increasing temperature. Among these materials, the interchain \(\kappa_{\mathrm{L}}\) values of GaSeI and AlSeI fall into the gray shaded region (below 0.1 W/mK) across a broad temperature range, which is widely recognized as the current limit for intrinsic \(\kappa_{\mathrm{L}}\) in inorganic crystalline solids. Specifically, the room-temperature \(\kappa_{\mathrm{L}}\) of GaSeI and AlSeI along the \(x\)-axis reach record-low values of 0.086 and 0.089 W/mK, respectively, and further drop to 0.058 and 0.059 W/mK at 900 K. This extreme thermal transport suppression indicates that the dual-channel thermal transport in these quasi-1D materials is synergistically suppressed, thereby pushing the thermal conductivity toward its fundamental minimum.

\section{Conclusions}\label{sec:3}
In summary, we demonstrate that strong phonon localization serves as an effective strategy to decouple the particle-like propagation (\(\kappa_{\mathrm{L}}^{\mathrm{P}}\)) and wave-like tunneling (\(\kappa_{\mathrm{L}}^{\mathrm{C}}\)) channels of heat transport, overcoming the competition that bottlenecks the lowest achievable lattice thermal conductivity in inorganic crystalline solids. The results reveal that, in the quasi-one-dimensional van der Waals (vdW) III–VI–VII ternary helical crystals, the weak interchain vdW interactions and heavy elements induce the emergence of numerous localized phonon branches. Consequently, both diagonal and off-diagonal components of the group velocity matrix are suppressed, thereby synergistically suppressing \(\kappa_{\mathrm{L}}^{\mathrm{P}}\) and \(\kappa_{\mathrm{L}}^{\mathrm{C}}\). Taking InSeI as an example, the average diagonal group velocity along the interchain direction is merely 0.073 km/s, nearly half of that along the intrachain direction (0.148 km/s), while the off-diagonal velocities are suppressed even more significantly, with values of 0.027 and 0.099 km/s for the two directions, respectively. Consequently, the interchain \(\kappa_{\mathrm{L}}^{\mathrm{P}}\) and \(\kappa_{\mathrm{L}}^{\mathrm{C}}\) at 300 K are only 0.145 and 0.053 W/mK, respectively, yielding an ultralow total \(\kappa_{\mathrm{L}}\) of 0.198 W/mK; the intrachain \(\kappa_{\mathrm{L}}\) remains a modest 0.466 W/mK. More remarkably, the interchain and intrachain \(\kappa_{\mathrm{L}}\) exhibit distinctly opposite mass dependencies. Along the intrachain direction, the lighter Ga- and Al-based compounds (GaSI, GaSeI, AlSeI) display significantly higher \(\kappa_{\mathrm{L}}\) than InSeI due to their stronger covalent bonding within the helical tubes. In contrast, these lighter systems exhibit lower \(\kappa_{\mathrm{L}}\) along the interchain direction, a direct consequence of the looser interchain packing and further weakened vdW interactions, as revealed by the more delocalized electron localization function and smaller interchain –ICOHP values. Therefore, the room-temperature \(\kappa_{\mathrm{L}}\) of GaSeI and AlSeI along the \(x\)-axis reach record-low values of 0.086 and 0.089 W/mK, respectively, and further drop to 0.058 and 0.059 W/mK at 900 K. Unlike conventional ultralow \(\kappa_{\mathrm{L}}\) materials dominated by coherent transport, these quasi-1D crystals exhibit a persistent negative temperature dependence, demonstrating that both transport channels are efficiently suppressed across a wide temperature range. These findings reveal dimensional-confinement-induced phonon localization as an effective materials design paradigm for simultaneously suppressing particle-like and coherent thermal conductivity, paving the way for the development of high‑performance thermal insulators. 

\section{Methods}\label{sec:4}
All density-functional theory (DFT) calculations are performed using the Vienna Ab initio Simulation Package (VASP)~\cite{RN68} with the projector augmented wave (PAW)~\cite{RN69} method. The exchange-correlation functional is treated within the generalized gradient approximation (GGA) using the Perdew–Burke–Ernzerhof (PBE) functional~\cite{RN70}, complemented by the DFT-D3 (IVDW=12) correction to account for van der Waals (vdW) interactions~\cite{RN71}. A plane-wave energy cutoff of \(500\) eV is adopted for all systems. Brillouin zone sampling is performed using the \(\Gamma\)-centered k-meshes, specifically 3×3×5 for InSeI, and 1×1×1 for GaSI, GaSeI, and AlSeI. Geometry optimizations are carried out until the residual forces on each atom fall below \(10^{-3}\) eV/Å and the total energy convergence reaches \(10^{-7}\) eV. Ab initio molecular dynamics (AIMD) simulations are performed in the NVT ensemble using a Nosé–Hoover thermostat with a 1.0 fs time step, and the systems are linearly heated from 10 to 1300 K over 20 ps. In this simulation, a 1×1×2 supercell (192 atoms) is constructed for InSeI, while the conventional cells (492 atoms) are directly employed for GaSI, GaSeI, and AlSeI. Subsequently, machine learning interatomic potentials (MLIPs) are constructed to efficiently simulate lattice dynamical properties. The temperature-dependent interatomic force constants (IFCs) for InSeI are extracted using the temperature-dependent effective potential (TDEP) method~\cite{RN40} from molecular dynamics trajectories driven by neuroevolution potential (NEP)~\cite{RN72}, in which interatomic forces are predicted by a high-precision multi-atomic cluster expansion (MACE) potential~\cite{RN73}. The interaction cutoff radii for the second-, third-, and fourth-order IFCs are set to 10.0, 6.5, and 4.0 Å, respectively. Additionally, harmonic second-order IFCs at 0 K are calculated using the finite-displacement method as implemented in the PHONOPY package~\cite{RN74}. The lattice thermal conductivity (\(\kappa_{\mathrm{L}}\)) is calculated by a unified thermal transport theory and implemented in the modified FOURPHONON software~\cite{RN22,RN75,RN76,RN77,RN78,RN79}. The equation for \(\kappa_{\mathrm{L}}\) is as follows:
\begin{equation}
	\begin{split}
		\kappa_{\mathrm{L}}^{\mathrm{P/C}} = \frac{\hbar^{2}}{k_{\mathrm{B}}T^{2}VN} \sum_{\mathbf{q}} \sum_{j,j^{\prime}} \frac{\omega_{\mathbf{q}j}+\omega_{\mathbf{q}j^{\prime}}}{2} \nu_{\mathbf{q}jj^{\prime}} \otimes \nu_{\mathbf{q}jj^{\prime}} \\
		\times \frac{\omega_{\mathbf{q}j}n_{\mathbf{q}j}(n_{\mathbf{q}j}+1) + \omega_{\mathbf{q}j^{\prime}}n_{\mathbf{q}j^{\prime}}\left(n_{\mathbf{q}j^{\prime}}+1\right)}{4\left(\omega_{\mathbf{q}j} - \omega_{\mathbf{q}j^{\prime}}\right)^{2} + \left(\Gamma_{\mathbf{q}j} + \Gamma_{\mathbf{q}j^{\prime}}\right)^{2}} \left(\Gamma_{\mathbf{q}j} + \Gamma_{\mathbf{q}j^{\prime}}\right),
	\end{split}
	\label{eq:eq-1}
\end{equation}
where the superscripts P and C represent the contributions from populations (particle-like propagation) and coherences (wave-like phonon tunneling), respectively. When \(j = j^{\prime}\), Eq.~\ref{eq:eq-1} reduces to the Peierls–Boltzmann transport equation (particle-like term); otherwise, it describes the wave-like interband tunneling (coherent term). The 10×10×20 N-grid is used to obtain converged \(\kappa_{\mathrm{L}}\) values. Moreover, the \(\kappa_{\mathrm{L}}\) is also obtained using homogeneous nonequilibrium molecular dynamics (HNEMD) simulations based on the trained NEP model and implemented in the GPUMD software~\cite{RN47,RN72,RN80}. The large supercells exceeding 90×90×90 Å$^{3}$ (containing over 20000 atoms) are adopted. The same supercell configuration is used to simulate the thermal expansion behavior. All computational details and convergence tests are provided in the Supporting Information (SI).

\section*{ACKNOWLEDGMENTS}
This work was supported by the National Natural Science Foundation of China (Nos. 52372260, 12574262, and 12174098), the Science Fund for Distinguished Young Scholars of Hunan Province of China (No. 2024JJ2048), the Hunan Provincial Innovation Foundation for Postgraduate (No. CX20250918), the Major Fundamental Research Program of Hunan Province (2025ZYJ004). The authors gratefully acknowledge the computing resources provided by the High-Performance Computing Platform of the School of Physics and Optoelectronics, Xiangtan University.

\section*{Competing interests}
The authors declare no conflict of interest.

\section*{Data Availability Statement}
The data supporting the findings of this study are available within this article and its Supporting Information. The trained NEP and MACE models are openly available from \url{https://github.com/ZhunyunTang/Q-1D}. Additional data are available from the corresponding author on reasonable request.

\section*{Code availability}
The source code for GPUMD is available at \url{https://github.com/brucefan1983/GPUMD}, for PHONOPY is available at \url{https://github.com/phonopy/phonopy}, for TDEP is available at \url{https://github.com/tdep-developers/tdep}, for MACE is available at \url{https://github.com/ACEsuit/mace}, for SHENGBTE is available at \url{https://www.shengbte.org}, for FOURPHONON is available at \url{https://github.com/FourPhonon/FourPhonon}, for the modified version of FOURPHONON is available at \url{https://github.com/ZhunyunTang/FourPhonon--wigner-sampling}.

\section*{Author contributions}
Z.T. contributed to the methodology, software, all calculations, formal analysis, visualization, and writing of original draft. X.W., J.L., and C.H. contributed to discussions and writing of manuscript. M.C., C.T. and T.O. were responsible for supervision, formal analysis, writing and discussion.

\section*{Keywords}
Phonon Localization, Dual-Channel Transport, Ultralow Lattice Thermal Conductivity, Quasi-One-Dimensional Materials.

\section*{References}
\bibliography{reference}
\end{document}